# Failure of Logarithmic Oscillators to Thermostat Small Atomic Clusters

Daniel Sponseller and Estela Blaisten-Barojas*

*Computational Materials Science Center and*

*School of Physics, Astronomy & Computational Sciences, George Mason University,*

*4400 University Drive, Fairfax, VA 22030-4422, USA*

(Dated: January 16, 2014)

## Abstract

A logarithmic oscillator has the outstanding property that the expectation value of its kinetic energy is constant for all stationary states [1]. Recently the ansatz that this property can be used to define a Hamiltonian thermostat has been put forward [2]. The latter publication suggests that the logarithmic oscillator weakly coupled to a small system would serve as a thermostat as long as few degrees of freedom are involved such as in atomic clusters. We have applied these ideas to a cluster of four Lennard-Jones atoms and inspected two different models of coupling between the cluster and the logarithmic oscillator in 3D. In both cases we show that there is a clear generation of kinetic motion of the cluster center of mass, but that kinetic energy due to interatomic vibrations are not significantly affected by coupling to the logarithmic oscillator. This is a failure of the published ansatz [2], as the logarithmic oscillator is unable to modify the kinetic energy due to vibrations in small atomic clusters.

*blaisten@gmu.edu; http://cmasc.gmu.edu

Computational thermostats are indispensable tools for scientists modeling and simulating cluster physics, molecules, biomaterials, and soft matter, among other systems that may involve a computationally reasonable number of degrees of freedom. Computational thermostats provide the means to produce controlled dynamics for systems by maintaining their kinetic energy at desired values [3]. Researchers in reference [2] suggested that a logarithmic oscillator (ln-oscillator) possessing the peculiar property of its average kinetic energy being a constant [1] may be used as a Hamiltonian thermostat to another system that is weakly coupled to it. The Hamiltonian of the ln-oscillator is

$$H_{ln} = \frac{p_x^2}{2\mu} + U_0 \ln \frac{|X|}{b}, \quad (1)$$

where $p_x$, $\mu$, and $X$ are the momentum, mass, and position, respectively. $U_0$ is the strength of the potential energy and $b$ is a positive length scaling factor. According to the virial theorem $<p \partial H_{ln} / \partial X> = <X \partial H_{ln} / \partial p>$ one obtains $<p^2/\mu>$, e.g. twice the expectation value of the kinetic energy is $U_0$ irrespective of the mass of the oscillator or of its energy. It is shown in reference [4] that the previous property implies infinite heat capacity making the ln-oscillator an ideal thermostat candidate by defining a kinetic temperature as $T = <p^2/\mu> = U_0$. The authors state that the most practical use of this ln-oscillator is as an analog thermostat for small systems, systems with a few degrees of freedom.

The Hamiltonian thermostat ansatz has been questioned [5, 6]. In addition, Hoover and Hoover [7] argue that the ability to establish heat flow in a system is a necessary test of a thermostat. If a proposed thermostat is incapable of transporting heat away from a hot reservoir to a cold reservoir such thermostat is not fit to control temperature. In order to analyze the virtues of the ln-oscillator these authors create a 1-D, 60-particle chain with 4-interacting particles. Twenty particles in each chain-end are connected to two different thermostats, one ln-oscillator is attached to each of the 20-particles on one end to simulate a thermostat and 20 other ln-oscillators are attached to the other chain end to account for the second thermostat. One chain-end is cold at T = 0.5 and the other chain-end is hot at T = 1.5. The 20 particles in the middle of the chain are equilibrated at T = 1.0 prior to beginning the simulation. The equations of motion are followed for some time to see if a linear temperature profile develops across the central chain length and a heat flux is established between the two thermostats in the steady state. Authors note that not only no heat flux was established but also the hot thermostated particles had temperatures far below the specified T = 1.5.

In this communication we describe two laboratory experiments that employ the hypotesized Hamiltonian thermostat by extending the ln-oscillator dimensionality to 3-D and coupling it weakly to a 4-atom Lennard-Jones cluster. In finite systems such as atomic clusters where the fluctuations of the density are on the order of the size of the system non-macroscopic thermodynamics enters into play [8]. We opt to avoid discussion about thermodynamics of the cluster and instead focus on how the time average of the cluster kinetic energy is affected when coupled to a ln-oscillator. In addition, along our simulations we were careful to select energies and parameters such that the atomic cluster remains bound without imposing external container walls. We considered two interaction models. In one the cluster is tethered to the origin and has a repulsive coupling to the ln-oscillator. In the second model the cluster and ln-oscillator are coupled through a week harmonic potential. The Hamilton equations of motion of the system were solved numerically and followed for 200,000,000 time steps. From these simulations we conclude that the ln-oscillator does not act as a bonafide thermostat able to change the time average kinetic energy of the atomic cluster. The main effect is a transfer of kinetic energy to the center of mass of the cluster. The kinetic energy of vibrations between atoms do not feel the presence of the ln-oscillator significantly. Therefore, our outcome is equally negative as the previous experiment of reference [7]. The following paragraphs describe in detail the two models used in this work and the results obtained.

Let us define a 3D ln-oscillator with a slightly modified potential energy:

$$H_{ln} = \frac{p^2}{2\mu} + U_0 \ln\left(\left(\frac{r}{b}\right)^2 + offset\right) \qquad (2)$$

where *r* is the distance of the ln-oscillator from the origin, *p* is its linear momentum, *b* is a length scaling parameter, and *offset* is a positive constant used to eliminate the singularity at *r = 0*. The mass of this oscillator is $\mu = 1$. The trajectory of this ln-oscillator is torus-shaped as shown in Figure 1. The virial theorem yields:

$$\left< \frac{p^2}{2\mu} \right> = U_0 \left< \frac{(r/b)^2}{(r/b)^2 + \text{offset}} \right> \quad (3)$$

showing that the average kinetic energy depends on the oscillator position and *offset*. For small values of offset, the average kinetic energy is very close to the constant $U_0$. In this work we used *offset = 0:0001* throughout, such that for $U_0 = 0.1$ the minimum of the potential at $r = 0$ is *-0.92103*.

Furthermore, let us consider a cluster of four Lennard-Jones atoms with mass $m = 1$ weakly coupled to one 3-D ln-oscillator. A sketch of the system is given in Figure 2. Two interaction models will be considered in which the cluster previously equilibrated and with its center of mass at rest is placed at the origin and the ln-oscillator is placed at a given distance from it. Model 1 has the cluster pinned to the origin by a restoring potential and linked to ln-oscillator by a repulsive potential. Model 2 has the cluster free floating with its center of mass at the origin and linked to the ln-oscillator through a parabolic potential with minimum at a distance *Roo*.

The Hamiltonian of model 1 is

$$H = \sum_{i=1}^{4} \frac{p_i^2}{2m} + 4\epsilon \sum_{i=1}^{3} \sum_{j>i}^{4} \left[ \left( \frac{\sigma}{R_{ij}} \right)^{12} - \left( \frac{\sigma}{R_{ij}} \right)^{6} \right] + H_{ln} + U_{int} + C \left( x_{cm}^4 + y_{cm}^4 + z_{cm}^4 \right) \quad (4)$$

$$U_{int} = \epsilon_{int} \sum_{i=1}^{4} \left( \frac{\sigma_{int}}{R_{io}} \right)^{12} \quad (5)$$

where $p_i$ and m are the linear momenta and masses of the four atoms, $R_{ij}$ are the distances between the respective atoms, $x_{cm}$, $y_{cm}$, and $z_{cm}$ are coordinates of the cluster's center of mass, $R_{io}$ are the distances between the ln-oscillator and each atom in the cluster, $\varepsilon_{int} = 0.036\varepsilon$, $\sigma_{int} = \sigma$, and $C = 1.25\ \varepsilon/\sigma^4$. The units adopted are the Lennard-Jones parameters $\varepsilon$, $\sigma$, $m$, and $\tau = \sqrt{m\sigma^2/\varepsilon}$ for energy, distance, mass, and time, respectively. The ln-oscillator is placed initially at $x = y = z = 3.5$ with initial velocities of $v_x = 0.08$, $v_y = 0.04$, and $v_z = 0.03$. The initial cluster configuration is a well-equilibrated cluster with average kinetic energy per particle <KE> = 0.1. The cluster is equilibrated for 10,000,000 time steps. The equations of motion are solved using the velocity-Verlet algorithm and a time step of 0.001 $\tau$. This method is symplectic. Simulations are run 200,000,000 time steps. Next the ln-oscillator is turned on by phasing in the value of the coupling strength in increments of 0.1, $\varepsilon_{int}$ every 10,000 time steps until reaching the full desired strength. We note that atoms evaporate from this small cluster when the average kinetic energy is about 0.25; thus, values selected for $U_0$ are below such threshold.

Figure 3 shows the kinetic (left pane) and potential (right pane) energies for three values of $U_0$. For $U_0 = 0.05$ the ln-oscillator should lower the kinetic energy of the cluster from 0.1 to the set value of 0.05. As seen in the top pane of Fig. 3, the kinetic energy of the ln-oscillator (asterisks) remains more or less at 0.05 while the kinetic energy per atom associated to vibrations (squares) makes excursions between its initial value of 0.1 and 0.05. Meanwhile a fair amount of kinetic energy is acquired by the cluster center of mass (black circles). Concerning the potential energy (right top pane of 3), the ln-oscillator (asterisks) and the cluster potential energy per atom (squares) are basically counterbalancing each other with an increase in the ln-oscillator and corresponding decrease in the cluster. The interaction energy (triangles) is slightly positive and small compared to the other system energies as expected. Increasing the constant $U_0 = 0.1$ (middle pane) has the ln-oscillator decreasing slowly its kinetic energy while generating a large loss of the cluster kinetic energy in the first half of the simulation time but maintaining about the original value in the second part of the simulation. Here again a significant kinetic energy is gained by the cluster center of mass as comparable strength than in the previous case. The potential energies of all parties behave similarly to the first case. For $U_0 = 0.2$ the kinetic energy of the ln-oscillator decreases to about 0.15 while the cluster kinetic energy per atom increases to 0.12 and the center of mass has a slightly larger and more constant value of 0.045 than in the previous two cases. Potential energies display a similar behavior as in the two previous cases. Figure 4a shows the approach between cluster and ln-oscillator over time and Figure 4b shows phase portraits of the ln-oscillator motion during 10,000,000 steps. After these observations we conclude that the ln-oscillator did not fulfill its task of thermostating the cluster. The most noticeable feature is a gain of kinetic energy by the center of mass of the cluster accompanied by minor changes in the kinetic energy due to vibrations. As shown in 4a the cluster and oscillator may be wildly separated over time depicting an undesirable behavior to maintain the vibrational kinetic energy controlled over time.

The Hamiltonian of model 2 is:

$$H = \sum_{i=1}^{4} \frac{p_i^2}{2m} + 4\epsilon \sum_{i=1}^{3} \sum_{j>i}^{4} \left[ \left(\frac{\sigma}{R_{ij}}\right)^{12} - \left(\frac{\sigma}{R_{ij}}\right)^{6} \right] + H_{ln} + U_{int} \qquad (6)$$

$$U_{int} = \epsilon_{int} \sum_{i=1}^{4} (R_{io} - R_{oo})^2 \qquad (7)$$

where $R_{io}$ are distances between cluster atoms and the ln-oscillator, $\varepsilon_{int} = 0.001\varepsilon$ and the parameter $R_{oo} = 6.0$ identifies the distance between cluster atoms and ln-oscillator for which the coupling between them vanishes. This interaction function ensures that the ln-oscillator 'floats' at about $R_{oo}$ from the cluster while weakly interacting with it. The strength of the interaction is comparable to that of model 1 but now the cluster is not pinned to the origin.

As in model 1, we run the simulation for 200,000,000 time steps for three values of $U_0$ and the results are shown in Fig. 5. As seen in the figure (left panes) for all three cases the ln-oscillator maintains its average kinetic energy, the cluster vibrational kinetic energy increases slightly with time, and the center of mass kinetic energy is larger as $U_0$ increases. The potential energy (right panes) of the ln-oscillator (asterisks) is positive for all three cases and increases with the value of $U_0$. The cluster potential energy is basically the same for all three values of $U_0$ and the interaction energy is very small. Figure 6a shows that the distance between ln-oscillator and cluster fluctuates a lot in this model but the two objects do not get apart as widely as in model 1. Figure 6b depicts phase portraits of the ln-oscillator during the first 10,000,000 time steps of the simulation. These diagrams show a more regular behavior than in model 1. Once again, the most prominent feature is that the cluster center of mass has gained kinetic energy, but unfortunately, the ln-oscillator fails once again to perform the task of a thermostat.

In conclusion, despite that ln-oscillators have the outstanding property of constant average kinetic energy irrespective of their mass or their total energy, they cannot be used in practice as thermostats of condensed systems with small number of degrees of freedom. Other computational experiments were carried out using a different interaction potential between the atoms in the cluster. For example, clusters with 4 and 5-rubidium atoms were also modeled [9]. The results are very similar to the ones presented in the previous paragraphs: systematically the ln-oscillator produces a transfer of kinetic energy to the cluster center of mass. This is peculiar. Applications of the property might result interesting in the study of small clusters confined to nanocavities or nanoporoses as a result of cavity breathing modes.

**Acknowledgment** We are grateful to Professor William G. Hoover for having brought to our attention the ansatz of ln-oscillators as computational thermostats for atomic clusters and for very interesting related exchange of ideas.

**Figure captions**

1. Torus shaped trajectory in coordinate space of the 3-D ln-oscillator with $U_0 = 0.1$, $b = 1$, offset = 0:0001, $\mu = 1$, and initial conditions: $x = y = z = 3.5$, $v_x = 0.08$, $v_y = 0.04$, $v_z = 0.03$.

2. Schematic representation of the Initial configuration of the ln-oscillator coupled to a 4-atom Lennard-Jones cluster.

3. Average kinetic and potential energies of the various components entering in model 1 for three values of $U_0$. Data points are coarse grained averages over 10,000,000 time steps. Squares depict the average cluster vibrational kinetic energy per atom in the left pane and the average potential energy per atom on the right pane. Asterisks depict the average kinetic energy (left pane) and average potential energy (right pane) of the ln-oscillator. The filled circles on the left panes are the average kinetic energy per atom of the cluster center of mass. Triangles on the right panes are the average interaction potential energy between the ln-oscillator and the cluster.

4. (a) Time behavior of the distance between ln-oscillator and cluster center of mass for model 1 with $U_0 = 0.1$; (b) Phase portraits of the ln-oscillator in model 1 during the first 10,000,000 time steps.

5. Average kinetic and potential energies of the various components entering in model 2 for three values of $U_0$. Data points are coarse grained averages over 10,000,000 time steps. Squares depict the average cluster vibrational kinetic energy per atom in the left pane and the average potential energy per atom on the right pane. Asterisks depict the average kinetic energy (left pane) and average potential energy (right pane) of the ln-oscillator. The filled circles on the left panes are the average kinetic energy per atom of the cluster center of mass. Triangles on the right panes are the average interaction potential energy between the ln-oscillator and the cluster.

6. (a) Time behavior of the distance between ln-oscillator and cluster center of mass for model 2 with $U_0$; (b) Phase portraits of the ln-oscillator in model 2 during the first 10,000,000 time steps.

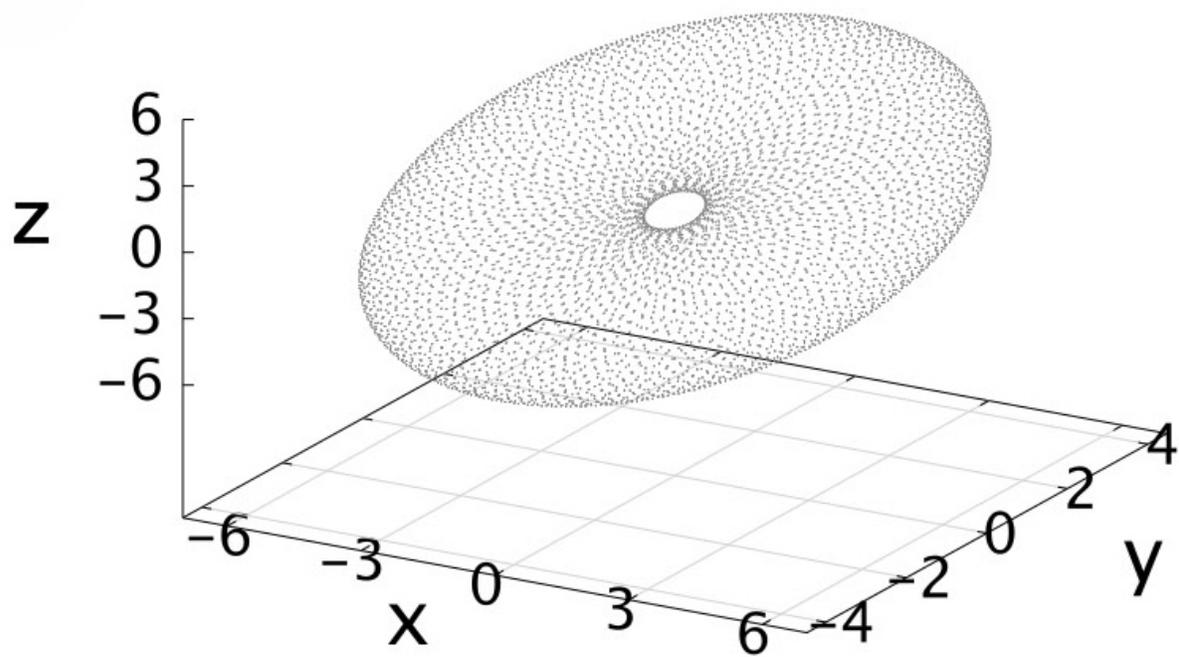

FIG 1

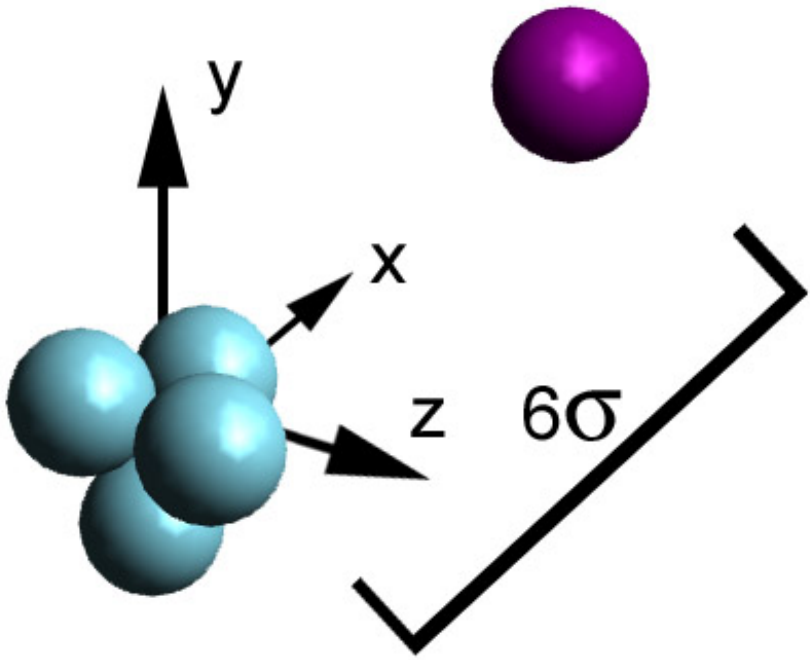

FIG. 2

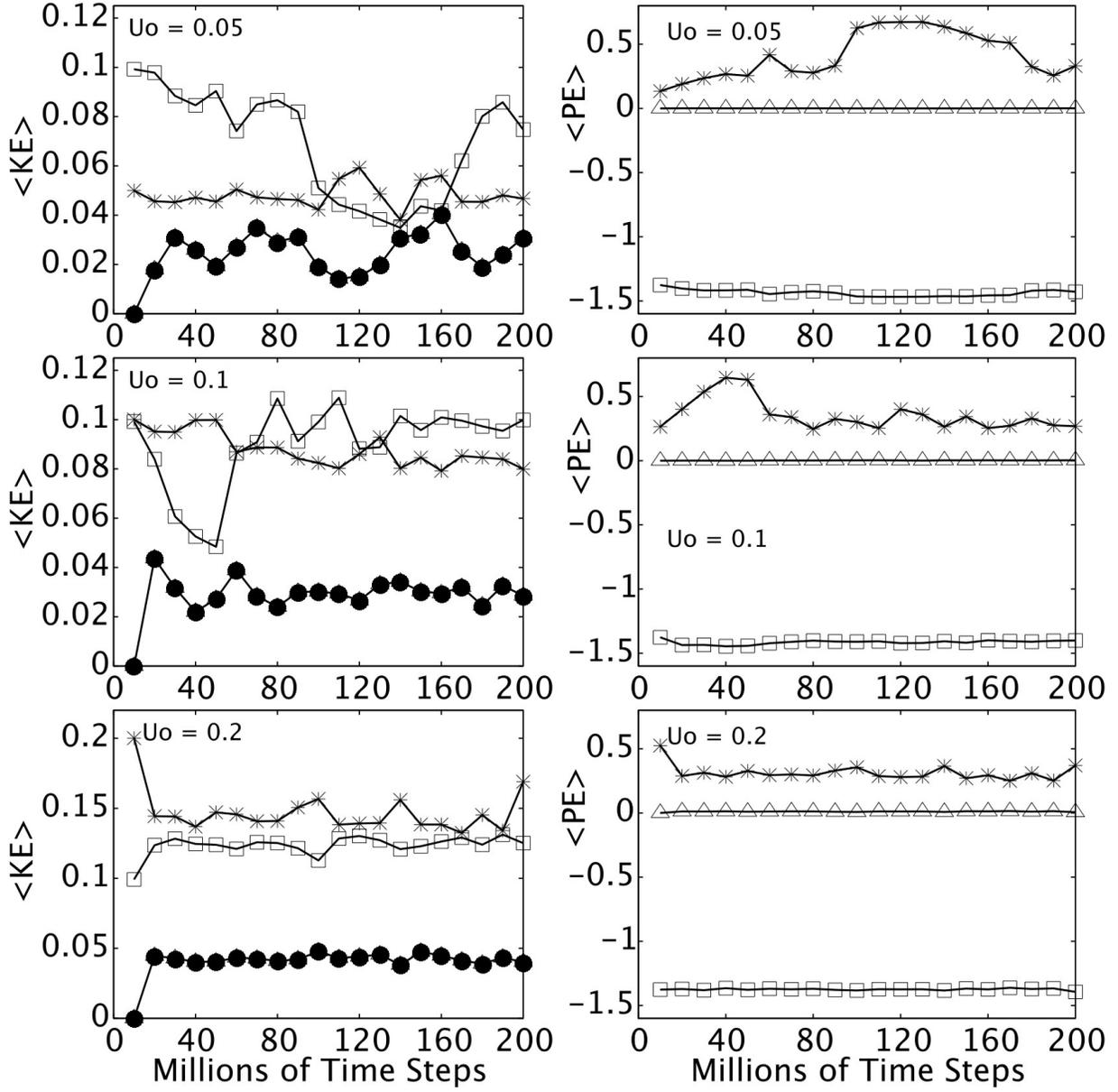

FIG. 3

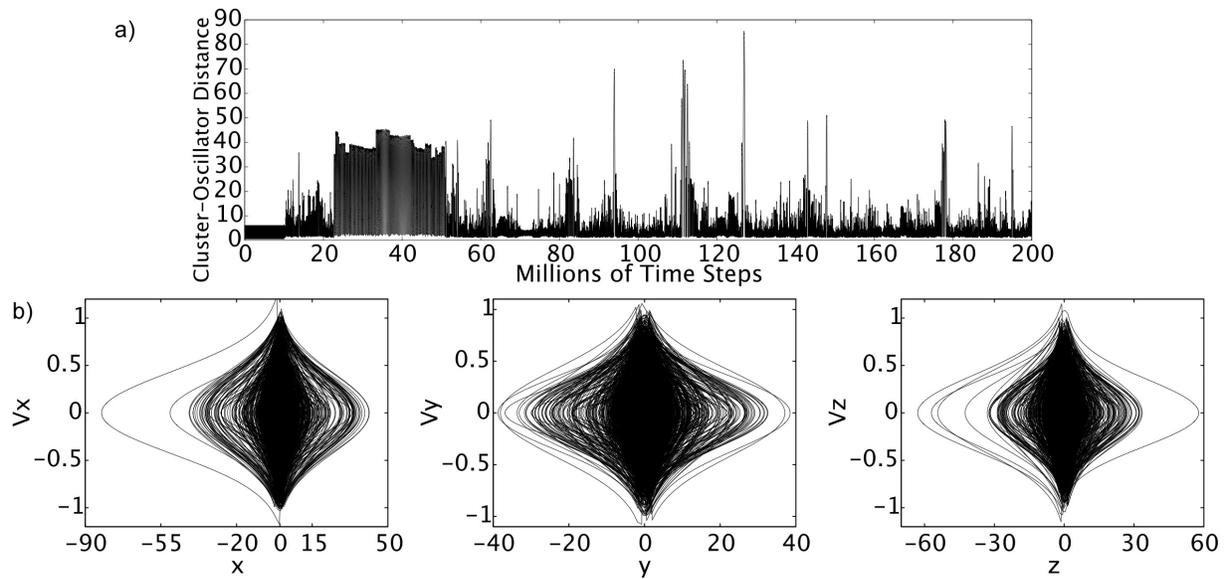

FIG. 4

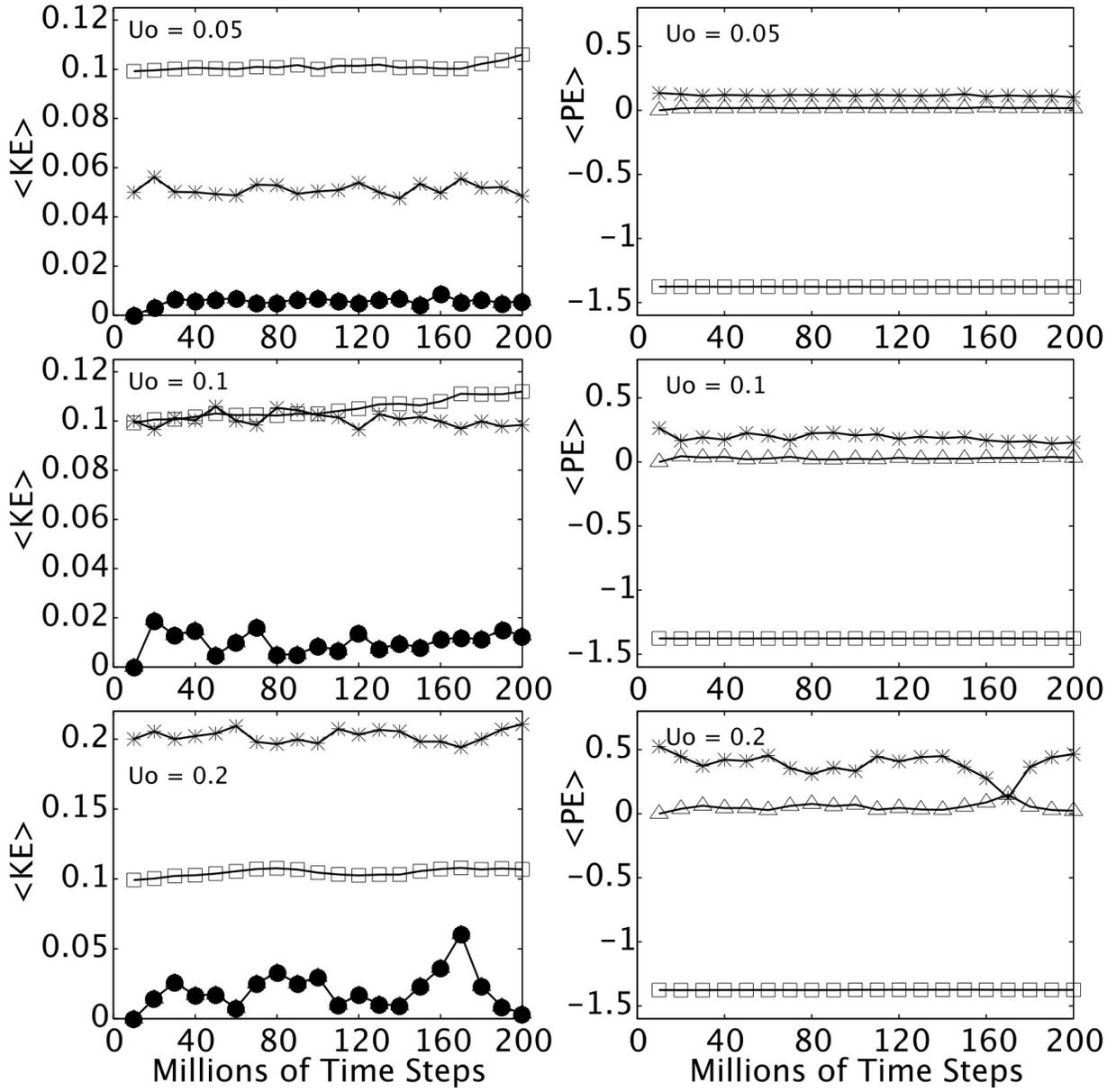

FIG. 5

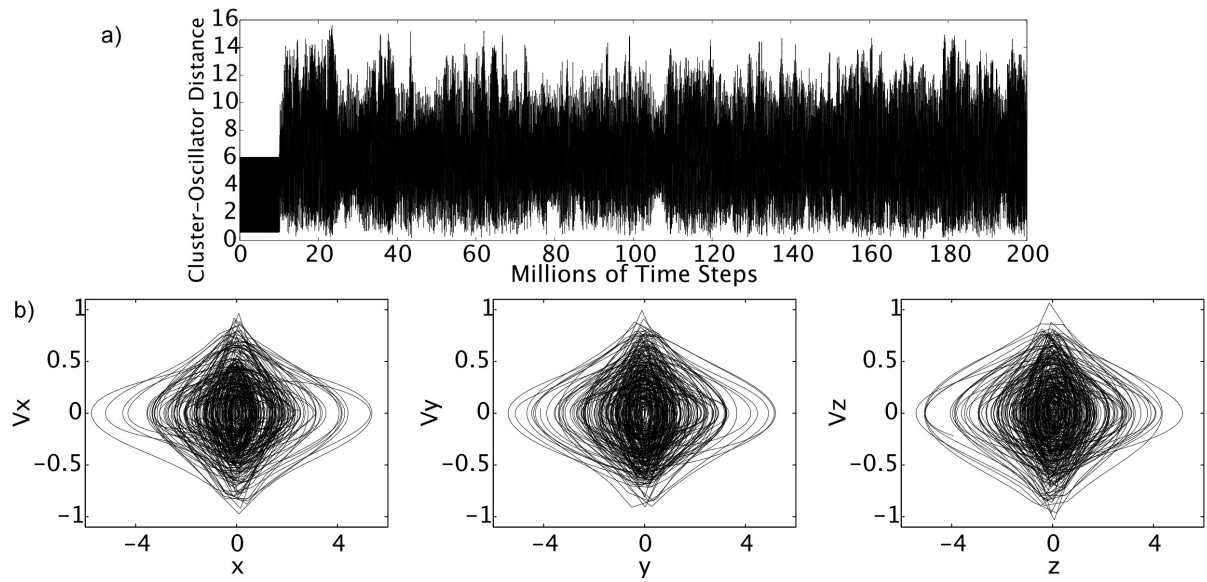

FIG 6